\documentclass[a4paper,11pt]{article}
\usepackage{jcappub} % for details on the use of the package, please
\usepackage[T1]{fontenc} % if needed

\usepackage{amssymb,amsmath,amsfonts}
\usepackage[dvipsnames]{xcolor}
\usepackage{graphicx}
\usepackage{longtable}
\usepackage{verbatim}
\usepackage{color}
\usepackage[mathscr]{euscript}
\usepackage{framed}
\usepackage{mdframed}
\usepackage{amsmath}
\usepackage{amsfonts}
\usepackage{mathrsfs}
\usepackage{amssymb}
\usepackage{mathrsfs}  
\usepackage{subcaption}
\usepackage{caption}
\usepackage{float}

\makeatletter
\renewcommand{\citation}[1]{%
  \g@addto@macro{\citation@list}{,#1}%
}
\newcommand*{\citation@list}{} % initialize
\newcommand{\sortbibitem}[2]{%
  \global\@namedef{bibitem@#1}{%
    \bibitem{#1} #2
  }%
}
\newcommand{\sort@bibitems}{%
  \@for\next:=\citation@list\do{%
    \@nameuse{bibitem@\next}%
    \global\@namedef{bibitem@\next}{}%
  }%
}
\expandafter\def\expandafter\endthebibliography\expandafter{%
  \expandafter\sort@bibitems\endthebibliography
}
\makeatother

\title{\boldmath A simple mechanism for the enhancement of the inflationary power spectrum}

\author{I. Dalianis$^{1}$,}
\author{A. Katsis$^{2}$}
\author{and N. Tetradis$^{2}$}
\affiliation{$^{1}$Department of Physics, University of Cyprus,
	Nicosia 1678, Cyprus 
	\\ $^{2}$National and Kapodistrian University of Athens,  Zographou 15784,  Greece}

\emailAdd{ntalianis.ioannis@ucy.ac.cy}
\emailAdd{ariskatsis@phys.uoa.gr}
\emailAdd{ntetrad@phys.uoa.gr}

\abstract{The background evolution in two-field inflation can feature two distinct stages, corresponding to the evolution along two 
	successive field directions. When the second stage occurs at a significantly lower energy scale, the inflationary trajectory 
	includes a sharp transition, accompanied by a series of rapid turns in field space. Fluctuations crossing the Hubble horizon during this turning phase can experience amplification by several orders of magnitude. This mechanism is very intuitive and can be
	implemented even in simple two-field models. It produces a peak in the scalar power spectrum that can lead to significant abundances of primordial black holes and secondary gravitational waves.}

\begin{document}
\maketitle
\flushbottom

\section{Introduction}\label{sec:intro}

Inflationary potentials with special structure can induce significant departures 
from scale invariance in the primordial spectrum 
of curvature perturbations ${\cal R}$. 
Such deviations may be confined to length scales far smaller than those directly 
constrained by cosmic microwave background (CMB) observations, thus leaving 
large-scale observables unaffected while enhancing power on small scales. 
This possibility is particularly important because perturbations generated 
during the later stages of inflation correspond to wavelengths well below those 
imprinted on the CMB sky and are therefore only weakly constrained by current 
cosmological data \cite{Planck:2018jri}. 
Nevertheless, a strong amplification of the curvature spectrum on such scales 
may have striking phenomenological consequences, including   the generation of a stochastic background of gravitational 
waves (GWs) sourced at second order by scalar perturbations 
\cite{Mollerach:2003nq,Ananda:2006af,Baumann:2007zm} and the formation of 
primordial black holes (PBHs) \cite{Hawking:1971ei,Carr:1974nx, Carr:2020gox, Carr:2025kdk}.
These signatures provide a powerful observational portal into inflationary 
physics far beyond the CMB window.

Multi-field inflation provides a natural and well-established framework
in which localized enhancement of ${\cal R}$ can arise dynamically.
Non-trivial trajectories in field space, such as bends or turns, are
generic in the presence of multiple scalar degrees of freedom and
potentials with curved directions
\cite{
Langlois:1999dw, Gordon:2000hv,
Cremonini:2010ua, Chen:2011zf,
Cespedes:2012hu, Achucarro:2012fd,
Konieczka:2014zja, Slosar:2019gvt,
Braglia:2020eai, Fumagalli:2020adf,
Fumagalli:2020nvq, Inomata:2021tpx,Greco:2025nvv}.
Whenever the inflationary trajectory undergoes a turn, isocurvature
(entropic) perturbations can efficiently source the adiabatic curvature
mode, leading to superhorizon evolution of ${\cal R}$
\cite{
Starobinsky:1994mh, Sasaki:1995aw,
Garcia-Bellido:1995hsq, Linde:1996gt,
Langlois:1999dw, Gordon:2000hv,
Achucarro:2010da, Cespedes:2012hu}.
A significant amplification takes place if during the sharp turn the
isocurvature modes experience a transition from heavy to light.
This sourcing mechanism opens up a wide range of possibilities for generating 
sharp features and localized peaks in the curvature power spectrum without 
violating observational constraints on large scales \cite{
Pi:2017gih, Palma:2020ejf,
Braglia:2020eai, Aldabergenov:2020bpt,
Fumagalli:2020adf, Fumagalli:2020nvq,
Kawai:2022emp, Christodoulidis:2023eiw,
Aldabergenov:2024fws, Wang:2024vfv,
Kim:2025dyi, GonzalezQuaglia:2025qem, Lorenzoni:2025kwn}.
Strong and localized enhancements of the curvature power spectrum can arise 
through the constructive superposition of multiple transient features in the 
background evolution. 
This phenomenon has been observed 
in the past in the context of models with scalar fields 
non-minimally coupled to gravity \cite{Pi:2017gih,He:2018gyf,Wang:2024vfv,Kim:2025dyi,Wang:2025dbj}.  

More recently, the constructive interference of multiple features was 
reanalysed in detail by the current authors, both in single-field inflation 
\cite{Kefala:2020xsx,Dalianis:2021iig} and in two-field inflation 
\cite{Boutivas:2022qtl}. (See also Ref. \cite{Dalianis:2023pur} for a concise review.)
%In particular, Ref.~\cite{Boutivas:2022qtl} showed that s
It was confirmed that sharp turns in field 
space can act as brief but efficient sources for the curvature perturbation, 
leading to resonant growth when several such events occur within a small 
number of efoldings.
Remarkably, this enhancement can take place even for canonical kinetic terms 
and a flat field-space metric, provided the trajectory experiences repeated 
bending. 
An explicit realization of this general mechanism was later constructed in 
Ref.~\cite{Katsis:2025eed}, where a simple class of two-field potentials was 
shown to exhibit successive turns along a sinusoidal valley. 
In this setup, a controlled interaction between the scalar fields dynamically 
carves out a turning trajectory in field space, effectively upgrading otherwise 
standard single-field models into two-field scenarios with additional 
phenomenological properties.
% \cite{Cespedes:2012hu,Katsis:2025eed}.
The enhancement of the curvature spectrum was characterized as
``assisted'' in Ref.~\cite{Katsis:2025eed} because it results through the influence of the 
isocurvature perturbations that arise temporarily during the
turns.

  \begin{figure}[t]
 \centering
\includegraphics[width=0.8\textwidth]{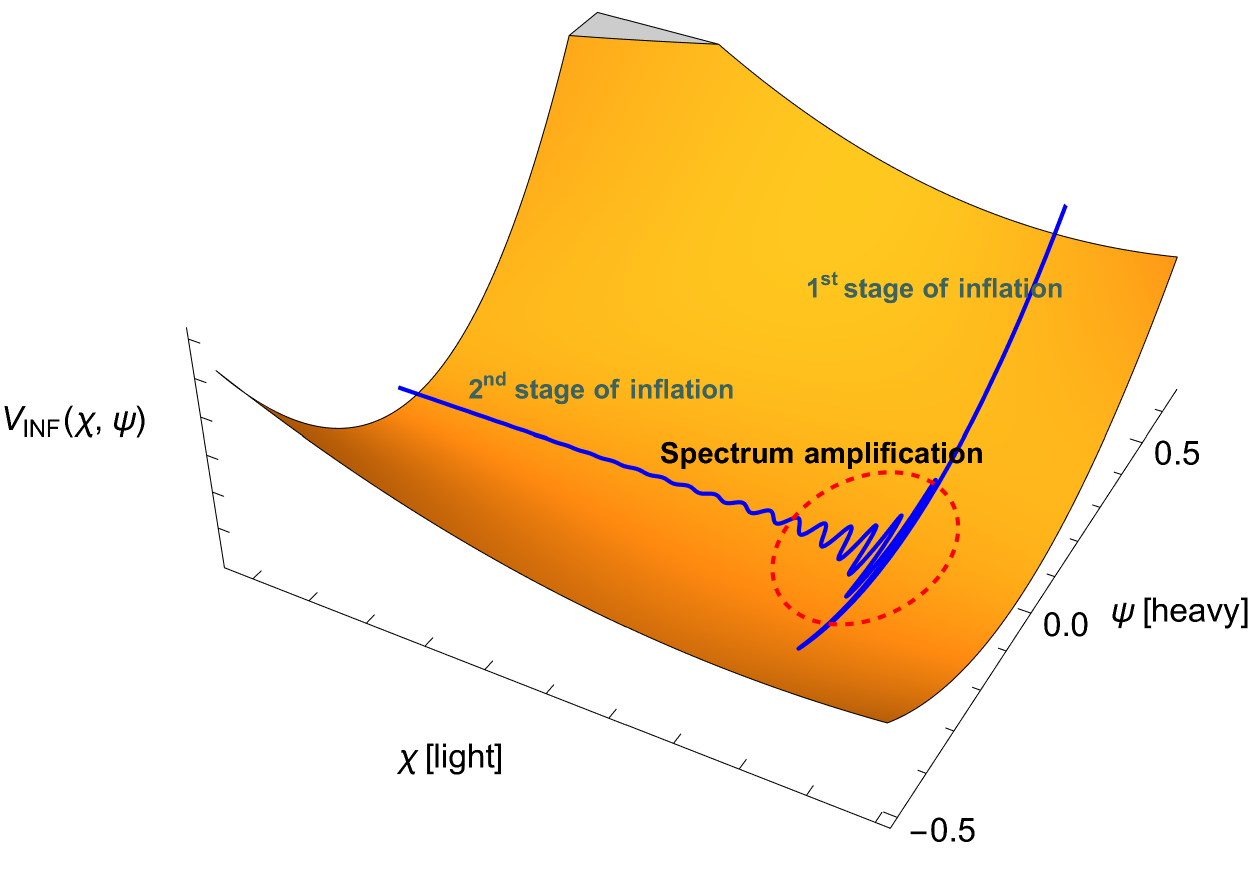}
\caption{
	A schematic illustration of the proposed minimal mechanism,  showing the qualitative shape of the inflationary potential and the corresponding field-space trajectory, with a zoom into the transition region.  The two stages of inflation are indicated in the plot.
}
\label{fig:intro}
\end{figure}

In this work, we argue that a qualitatively similar behaviour can emerge even in the absence of finely tuned or explicitly engineered features in the 
inflationary potential. 
We introduce a very simple bottom-up construction that makes use of 
the effect of turns in field space on the spectrum of adiabatic perturbations. The constructive interference of multiple turns can be efficient enough to account for the several orders of magnitude of enhancement desirable for model-building.
As we will demonstrate, the geometry of field-space trajectories in our
simple two-field models can naturally mimic the enhancement mechanisms 
previously associated with more elaborate constructions.
The key ingredients are an intermediate oscillatory regime 
and a significant separation between the energy scales of the two inflationary plateaux. This observation considerably broadens the class of inflationary models capable of generating large small-scale curvature perturbations.

We present the underlying mechanism within a minimal two-field inflationary 
setup featuring canonical kinetic terms and a simple interaction potential. 
The inflationary dynamics unfolds in two stages: an initial phase at a higher 
energy scale that fixes the CMB normalization, followed by a second stage at a 
lower energy scale. 
As the system transitions between these stages, the heavier field undergoes 
damped oscillations, which induce repeated turns in the background trajectory 
in field space. 
These turns act as transient sources for the curvature perturbation, leading 
to a localized amplification of the power spectrum over a narrow range of 
scales. An illustration of this mechanism is displayed in Fig. \ref{fig:intro}.

The resulting enhancement has direct and testable implications for PBH  formation, the generation of scalar-induced gravitational waves (SIGWs), and  the appearance of non-Gaussian signatures at the relevant small scales. 
In particular, the associated SIGW background can fall within the sensitivity 
ranges of pulsar timing array experiments as well as future space-based 
gravitational wave interferometers.

We discuss the mechanism of ``assisted enhancement'' in 
Section~\ref{sec:2}. We summarize the basic formalism for
two-field inflation in 
Subsection~\ref{sec:equ} and study a simple mode that realizes
the mechanism in Subsection~\ref{sec:cliff1}. 
We analyse the dependence of the power spectrum on the parameters of the model that
control the evolution of the 
background and fluctuations in Subsection~\ref{sec:params}.
We comment on how this 
mechanism may appear in more general inflationary scenarios in 
Section~\ref{sec:gen}. 
The phenomenological implications for GWs and PBHs are studied in Section~\ref{sec:PBHGW}. 
We present our conclusion in Section~\ref{conclusions}.
Throughout the paper we express all dimensionful quantities in units of $M_{\rm Pl}$, 
suppressing factors of $M_{\rm Pl}$ for notational simplicity.

\section{Assisted enhancement from the transition between two inflationary plateaux} 
\label{sec:2}

\subsection{The inflationary evolution of the background and fluctuations}\label{sec:equ}

We consider a simple action of two scalar fields, $\vec{\phi} =(\chi,\psi)$, with canonical kinetic terms, interacting through the potential $V(\chi, \psi)$:

\begin{equation}\label{equ:action}
S= \int d^4 x \sqrt{-g}\left[\frac{R}{2}-\frac{1}{2}g^{\mu \nu} \partial_{\mu} \chi \partial_{\nu}\chi-\frac{1}{2}g^{\mu \nu} \partial_{\mu} \psi \partial_{\nu}\psi  -V(\chi,\psi)\right]  \ .
\end{equation}
in a flat FRW metric. We measure time in terms of efoldings of expansion. We also rotate the field-space axes and parametrize with respect to the directions tangent and normal to the trajectory in field space \cite{Gordon:2000hv}.
The basic elements of the inflationary evolution in our notation are already described in \cite{Cespedes:2012hu,Achucarro:2010da,Katsis:2025eed}. 
The  background evolves as

\begin{subequations}\label{equ:eom}
\begin{align}
\label{equ:eom:1}
\chi_{,NN}+\left(3+\frac{H_{,N}}{H}\right)\chi_{,N}+\frac{V_{,\chi}}{H^2} & = 0 
\\
\label{equ:eom:2}
\psi_{,NN}+\left(3+\frac{H_{,N}}{H}\right)\psi_{,N}+\frac{V_{,\psi}}{H^2} &= 0 
\\
\label{equ:eom:3}
\frac{2V}{6-(\chi_{,N}^2+\psi_{,N}^2)} &= H^2 \ .
\end{align}
\end{subequations}
The parameter $\eta$ is also decomposed into its projections along these newly defined orientations

\begin{equation}\label{equ:eps}
\epsilon = \frac{1}{2}(\chi^2_{,N}+\psi^2_{,N}) \ ,
\end{equation}
\begin{equation}
\eta_{\parallel} = 3+\frac{V_{,\chi} \chi_{,N}+V_{,\psi}  \psi_{,N}}{H^2(\chi^2_{,N}+\psi^2_{,N})} \ ,
\end{equation}
\begin{equation}\label{equ:etan}
\eta_{\perp} = \frac{V_{,\chi} \psi_{,N}-V_{,\psi} \chi_{,N}}{H^2(\chi^2_{,N}+\psi^2_{,N})} \ .
\end{equation}
The parameter $\eta_{\parallel}$ is equivalent to the standard second slow-roll parameter of single-field inflation, while $\eta_{\perp}$ parametrizes the turning rate of the trajectory in field space. 

We describe fluctuations from the classical trajectory in terms of the curvature and isocurvature fields, $\mathcal{R}$ and $\mathcal{F}= \sqrt{2 \epsilon} \mathcal{S}$ respectively \cite{Gordon:2000hv,Achucarro:2010da,Cespedes:2012hu,Achucarro:2012sm,Polarski:1994rz,Mukhanov:1997fw}. The spectrum of the fluctuations is determined by the Mukhanov-Sasaki coupled evolution

\begin{subequations}\label{equ:flu}
\begin{align}
\label{equ:flu:1}
\mathcal{R}_{k,NN}+r_1(N) \mathcal{R}_{k,N}+r_2(N)k^2\mathcal{R}_{k} &= -r_3(N)\mathcal{F}_{k,N}-r_4(N)\mathcal{F}_k \ ,
\\
\label{equ:flu:2}
\mathcal{F}_{k,NN}+s_1(N) \mathcal{F}_{k,N}+\left(s_4(N)+r_2(N)k^2\right)\mathcal{F}_{k}&= s_3(N)\mathcal{R}_{k,N}  \ ,
\end{align}
\end{subequations}
where $\mathcal{R}_k$ and $\mathcal{F}_k$ are Fourier-transformed quantities and everything is expressed in terms of efoldings $N$. 
The power distributed to each mode can be deduced by squaring the amplitude of the corresponding fluctuation:
\begin{subequations}\label{equ:PS}
\begin{align}
\mathcal{P}_{\mathcal{R}}(k) &= \frac{k^3}{2 \pi^2}  |\mathcal{R}_k |^2 \ ,
\\
\mathcal{P}_{\mathcal{F}} (k)&= \frac{k^3}{2 \pi^2}  |\mathcal{F}_k |^2  \ .
\end{align}
\end{subequations}
The explicit formulas for the evolution coefficients are:
\begin{subequations}
\begin{align}
r_1 \equiv 3+\epsilon-2 \eta_{\parallel} \ , \quad r_2 \equiv \frac{e^{-2N}}{H^2} \ , \quad r_3 &\equiv 2 \frac{\eta_{\perp}}{ \sqrt{2 \epsilon}} \ , \quad r_4 \equiv 2 \frac{\eta_{\perp}}{\sqrt{2 \epsilon}}\left(3-\eta_{\parallel}+\frac{\eta_{\perp,N}}{\eta_{\perp}}\right)
\\
s_1  \equiv 3-\epsilon \ , \quad s_3 &\equiv 2  \sqrt{2 \epsilon} \eta_{\perp} \ , \quad s_4 \equiv   (\mu/H)^2 - \eta_{\perp}  \  ,
\end{align}
\end{subequations}
A useful quantity that we employ as a background-level precursor of enhancement is
\begin{equation}\label{equ:cond}
W \equiv   (\eta_{\perp})^2- (\mu/H)^2=-(M_{\rm eff}/H)^2 \ ,
\end{equation}
where  $\mu$ is the bare mass of the fluctuation orthogonal to the background trajectory, given by 
\begin{equation}\label{equ:M}
\mu^2 \equiv N_{\chi}N_{\chi}V_{, \chi \chi} +N_{\psi}N_{\psi}V_{, \psi \psi}+2N_{\chi}N_{\psi}V_{, \chi \psi} \ ,
\end{equation}
with
\begin{equation}
N_{\chi}=\frac{\dot{\psi}}{\sqrt{\dot\chi^2+\dot\psi^2}} \ , \quad N_{\psi}=-\frac{\dot{\chi}}{\sqrt{\dot\chi^2+\dot\psi^2}} \  .
\end{equation}
The parameter $M^2_{\rm eff}$ corresponds to the coefficient of the $\mathcal{F}^2$ term in the expansion of the action, which
can be interpreted as the effective mass term of the fluctuation \cite{Cespedes:2012hu}.

The mechanism of assisted enhancement can be summarized as follows\footnote{See \cite{Katsis:2025eed} for a more detailed analysis.}: The bending of the trajectory is encoded in the turning parameter $\eta_{\perp}$ and is determined by the background evolution. During the bending, a pulse of positive $W$ is generated that transiently drives $M_{\rm eff}^2$ tachyonic, leading to the exponential growth of isocurvature perturbations. The adiabatic mode responds to this increase because it is coupled to the isocurvature one through the turning parameter \cite{Garcia-Saenz:2018ifx}. Later, when the turn or series of turns is completed, the isocurvature mode decays but the amplitude of the adiabatic mode freezes at an increased value.

\subsection{A simple scenario of assisted
	 enhancement}\label{sec:cliff1}

 \begin{figure}[t]
 \centering
\includegraphics[width=0.6\textwidth]{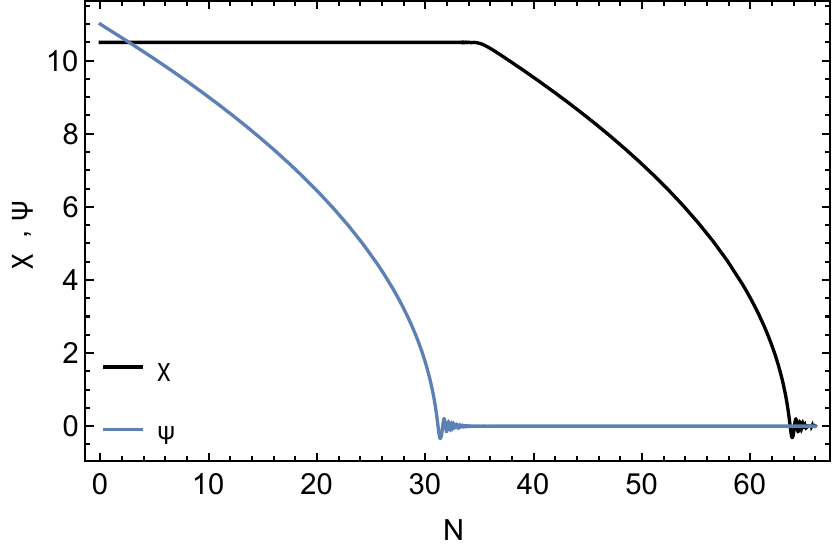}
\caption{The background solutions for $\chi(N)$ (black) and $\psi(N)$ (blue) for the model \eqref{equ:xsq} with parameters $m_{\chi}^2=8 \times 10^{-12}$, $m_{\psi}^2=4 \times 10^{-7}$ and $c_w=4 \times 10^{-3}.$
}
\label{fig:evo}
\end{figure}

\begin{figure}[t]
  \centering
  \begin{subfigure}[t]{0.58\textwidth}
    \centering
    \includegraphics[width=\textwidth]{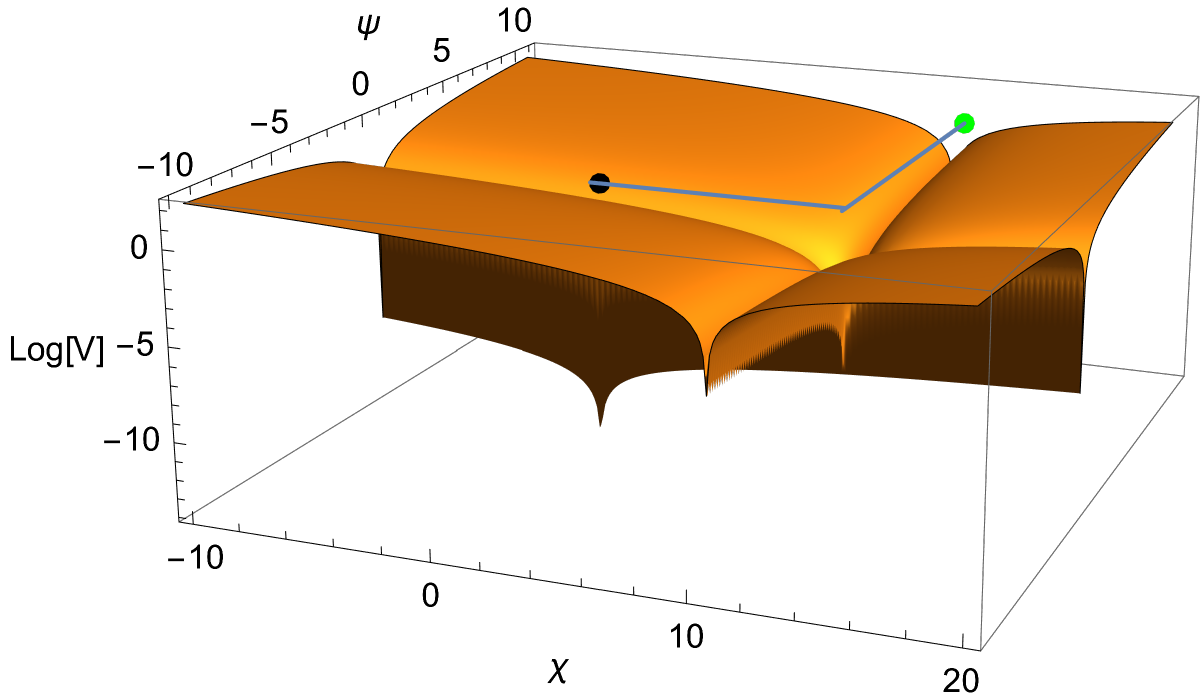}
  \end{subfigure}
  \hfill
  \begin{subfigure}[t]{0.35\textwidth}
    \centering
    \includegraphics[width=\textwidth]{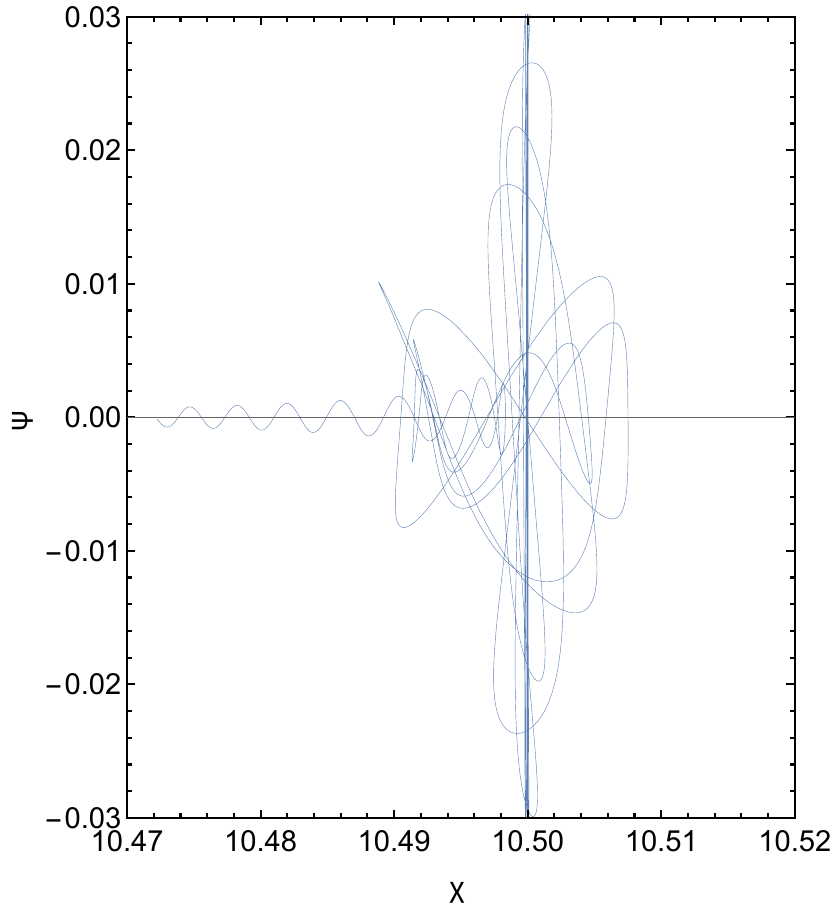}
  \end{subfigure}
  \caption{Left panel: The field trajectory shown on a logarithmic scale for the potential \eqref{equ:xsq}, highlighting its evolution across different energy scales. 
Right panel:  A zoomed-in view of the trajectory in
 field space during the transition stage, highlighting the multiple turns that characterize the two-field dynamics.
Both axes are in Planck units. }
  \label{fig:traj}
\end{figure}

\begin{figure}[t]
\centering
\includegraphics[width=0.50\textwidth]{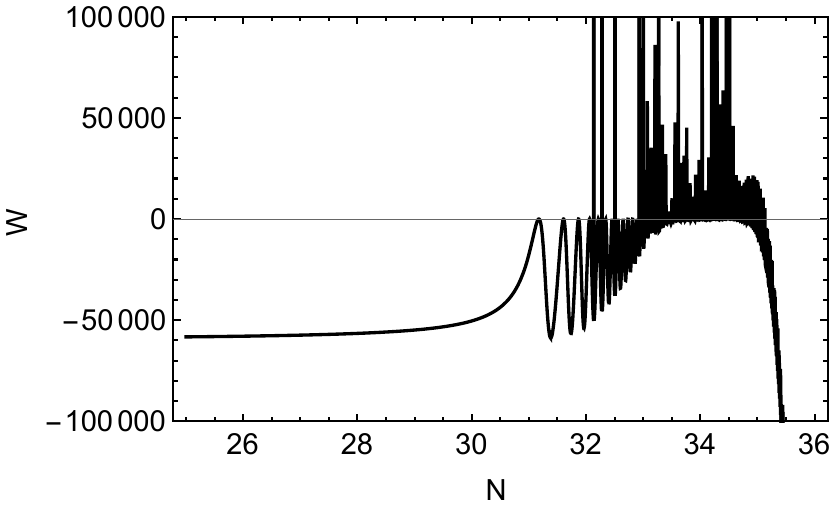}
\includegraphics[width=0.48\textwidth]{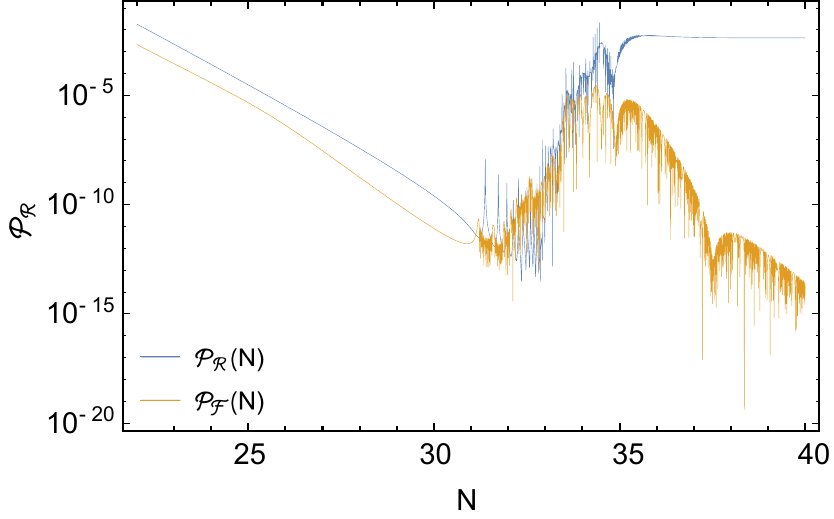}\,\,\,
\caption{ { Left panel}:
The evolution of the parameter $W$ defined in \eqref{equ:cond}. It is negative during the first stage of evolution, while large positive pulses occur during the transition stage.
{ Right panel}:  For the Fourier mode $k_{\rm p} = 7.6 \times 10^{10} \  {\rm Mpc}^{-1}$ we plot $\mathcal{P}_{\mathcal{R}}(N)$ (blue) and $\mathcal{P}_{\mathcal{F}}(N)$ (yellow) for the same model.
 }
\label{fig:Qfl1}
\end{figure}

\begin{figure}[t]
  \centering
  \begin{subfigure}[t]{0.49\textwidth}
    \centering
    \includegraphics[width=\textwidth]{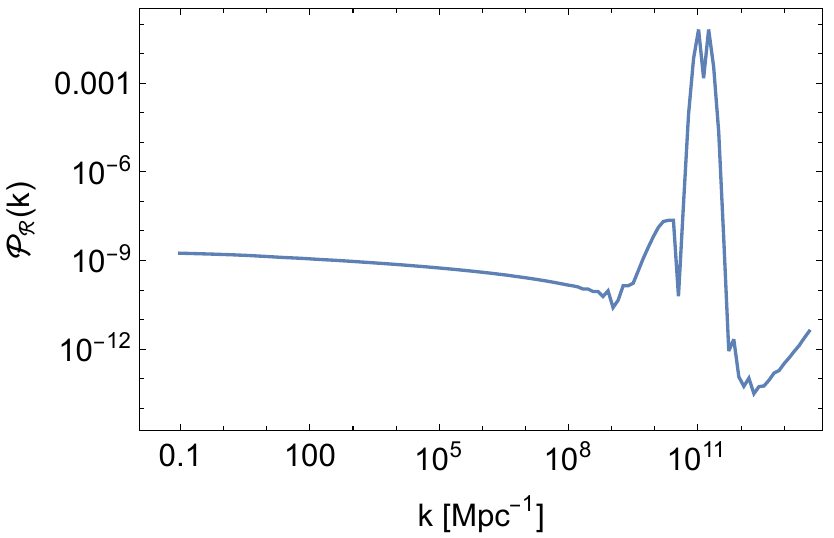}
  \end{subfigure}
  \hfill
  \begin{subfigure}[t]{0.49\textwidth}
    \centering
    \includegraphics[width=\textwidth]{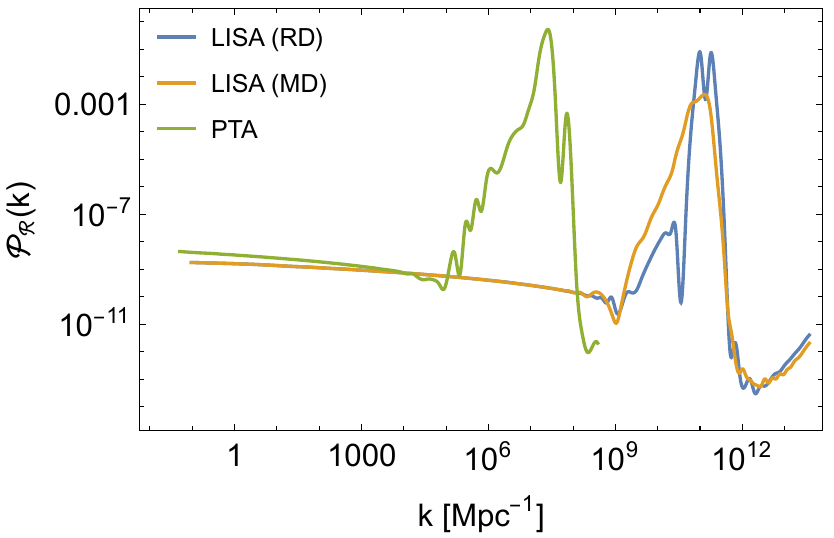}
  \end{subfigure}
  \caption{{ Left panel}: A representative shape of the scalar power  spectrum.
{ Right panel}: Three power spectra whose phenomenological implications for GWs and PBHs  are discussed in Section
\ref{sec:PBHGW}.
  }
 \label{fig:PRs}
\end{figure}

When considering the background evolution equations \eqref{equ:eom} it becomes apparent that the problem is analogous to a ball travelling on a two-dimensional $(\chi,\psi)$ terrain with height $V(\chi,\psi)$ and friction encoded in $H$. The crucial observation is that one does not have to carve a snaking valley in the potential in order to generate a trajectory with successive turns. Even for the inside of a tilted cylinder, if we place the ball at some random position away from the tilted valley which we take in the $\chi$ direction, the ball will naturally oscillate (with decaying amplitude) in the perpendicular $\psi$ direction,
while also moving down the valley. 
This type of evolution can take place in two-field models 
at the end of inflation before or during reheating. On the other
hand, we argue here that it can also arise during the 
stage of evolution between two inflationary plateaux, when
the vacuum energy is substantial. If the shape of the potential
is such that the turns in field space are sharp,
the spectrum of fluctuations around the background trajectory 
can be strongly enhanced.

In order to confirm that this scenario does not require elaborate model building, we consider here a simple potential that will serve as our reference model throughout the rest of the paper:
\begin{equation}\label{equ:xsq}
V(\chi,\psi)= m_{\chi}^2 \chi^2 +  m_{\psi}^2 \psi^2 
+ c_w \psi^2 (\chi-\chi_0)^2 \ .
\end{equation}
This potential features two valleys, located at $\psi=0$ and $\chi=\chi_0$. 
If the system initially starts with $|\psi|>0$, the field $\psi$ slowly rolls 
along the valley near $\chi=\chi_0$. During this first stage, the inflationary 
energy density is dominated by the contribution $m_\psi^2\psi^2$.
As $\psi$ approaches the lowest point of the first valley, it begins to oscillate around
 $\psi=0$. At the same time the field 
configuration gradually transitions into the second valley at $\psi=0$,  where the second and final stage of inflation takes place with the energy density dominated by the term $m_\chi^2 \chi^2$. A schematic presentation of the field evolution is 
given in Fig. \ref{fig:intro}.

For a more precise discussion, we focus 
on an explicit example with benchmark parameters. The masses are chosen as $m_{\chi}^2 = 8 \times 10^{-12}$ and $m_{\psi}^2 = 4 \times 10^{-7}$, with an interaction constant $c_w = 4 \times 10^{-3}$. The evolution starts at $N = 0$ from the initial field values $(\chi_0, \psi_0) = (10.5, 11)$. Figure~\ref{fig:evo} shows the resulting background evolution for this benchmark configuration.
It consists of two inflationary stages, approximately one in each
field direction, and a brief intermediate phase during which inflation is interrupted. During the first stage the heavier field $\psi$ makes its descent towards $\psi=0$, while  the lighter field $\chi$ is  frozen at the value $\chi=\chi_0$. When $\psi$ approaches its minimum value $\psi=0$
at an intermediate time $N\simeq 31$, it begins to oscillate about this point until 
approximately $N\simeq 36$. This short interval of a few efoldings constitutes 
the transition stage. When the amplitude of the $\psi$ oscillations
drops sufficiently, the system 
enters the second stage of inflation, during which $\chi$ slowly rolls toward 
the absolute minimum of the potential. 
The true vacuum, where reheating is expected to take place, is reached at 
$N\simeq 65$ and corresponds to $(\chi,\psi)=(0,0)$. For different choices of 
model parameters, the duration of inflation and the corresponding efoldings
can vary, allowing for a smaller or larger total value of $N$ to be realized.
We also point out that an additional initial stage is possible
if the system does not start exactly on the first valley of the
potential. This is brief, as the effective friction quickly 
forces the system into a slow-roll evolution along the
first valley.

\begin{table}[H]
\centering
\begin{tabular}{lccccc}
\hline\hline
GW probe
& $m_\chi^2$ 
& $m_\psi^2$ 
& $c_w$ 
& $\chi_0$ 
& $\psi_0$ \\
\hline
PTA 
& $3.2\times10^{-11}$ 
& $1.0\times10^{-6}$ 
& $0.01$ 
& $12.0$ 
& $9.3$ \\

LISA (RD) 
& $8.0\times10^{-12}$ 
& $4.0\times10^{-7}$ 
& $0.004$ 
& $10.5$ 
& $11.0$ \\

LISA (MD) 
& $3.16\times10^{-11}$ 
& $4.0\times10^{-7}$ 
& $0.004$ 
& $8.0$ 
& $11.0$ \\
\hline\hline
\end{tabular}
\caption{Model parameters,  in Planck units, for the potential \eqref{equ:xsq} and scenarios relevant for the PTA and LISA frequency bands.}
\label{tab:params}
\end{table}

In the left panel  of Fig. \ref{fig:traj} we show the potential on a logarithmic scale in order to make the slopes visible, with the field trajectory overlaid. A zoomed-in view of the trajectory during the transition stage is shown in the right panel.
 For $31<N<33$ there are large oscillations in the $\psi$ axis while $\chi$ remains at $\chi_0$. For $33<N<34$, strong turns in field space take place that display a similarity with Lissajous curves. For $34<N<35$, the trajectory moves from $\chi_0$ towards the minimum in a sine-type curve, involving many smaller turns in field space. In the left panel of Fig. \ref{fig:Qfl1} we depict the parameter $W$ defined in \eqref{equ:cond}. It is strictly negative during the first stage. On the other hand, the oscillations during the transition between the two stages produce a series of pulses of positive $W$. These pulses render the isocurvature fluctuations unstable.
In the right panel of Fig. \ref{fig:Qfl1} we show the evolution of the amplitude of the isocurvature (yellow) and adiabatic (blue) fluctuations for $k_{\rm p} = 7.6 \times 10^{10} \  {\rm Mpc}^{-1}$. When the pulses of positive $W$ occur, both isocurvature and adiabatic amplitudes increase in a series of small steps. Then the isocurvature amplitude decays and the adiabatic amplitude freezes at a value increased by $7$ orders of magnitude.
In the left panel of Fig.~\ref{fig:PRs} we present the numerical results for the curvature power spectrum $\mathcal{P}_{\mathcal{R}}$.
A relatively narrow peak appears around $k_{\rm p}$, while the amplitude  remains at the familiar $A_s$ value around $k_*= 0.05\, {\rm Mpc}^{-1}$. 
 In the right panel we present this spectrum together with two additional ones, 
corresponding to parameter choices different from the benchmark configuration.
The corresponding parameter values are listed in Table~\ref{tab:params}.  Their phenomenological implications are  discussed in Section
\ref{sec:PBHGW} and are illustrated in Fig. \ref{fig:GWPBH}.

\subsection{The dependence on the parameters of the model}\label{sec:params}

\begin{figure}[t]
\centering
\includegraphics[width=0.48\textwidth]{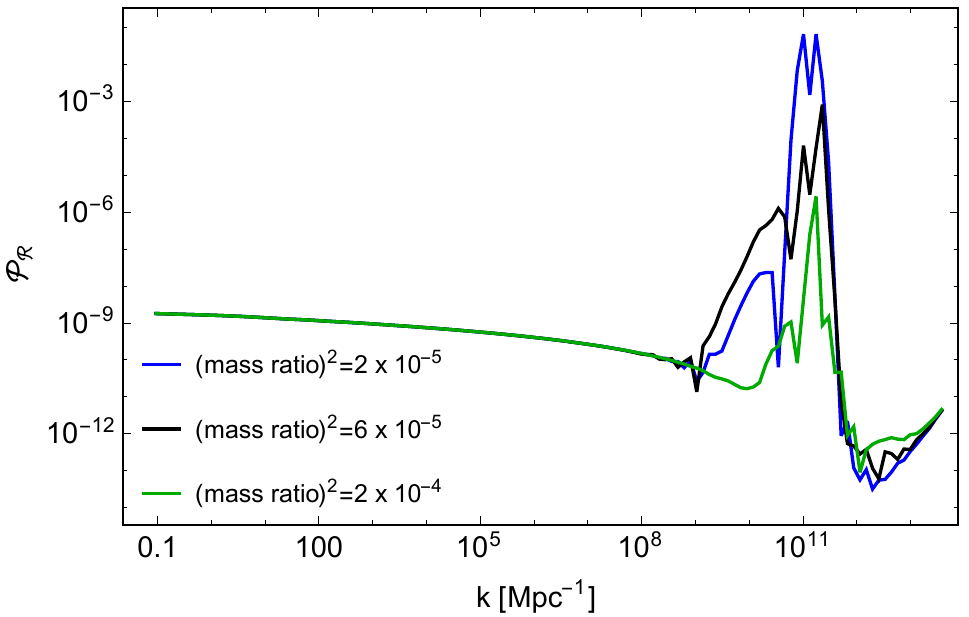}\,
\includegraphics[width=0.48\textwidth]{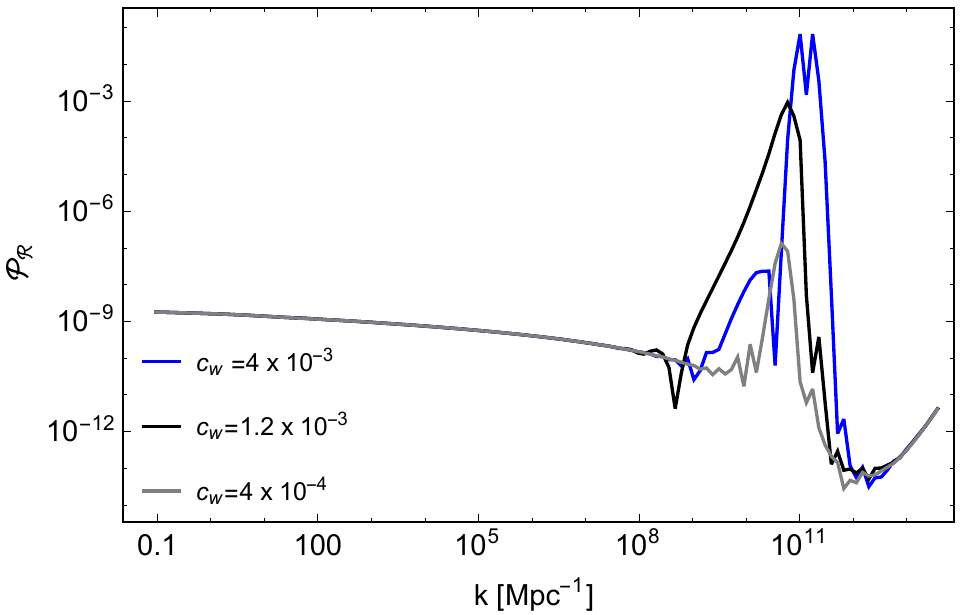}
\caption{The scalar power spectrum of the adiabatic perturbation, $\mathcal{P}_{\mathcal{R}} (k)$, for the band of Fourier modes $0.01\, {\rm Mpc}^{-1} \leq k \leq 10^{14}\, {\rm Mpc}^{-1}$ at $N=40$. In each panel we depict its variation as we change one parameter. Left: mass ratio $(m_{\chi}/m_{\psi})^2 = 2 \times 10^{-5}$ (blue),  $2 \times 10^{-4}$  (green), $6 \times 10^{-5}$ (black). Right: interaction strength  $c_w = 4 \times 10^{-3}$ (blue),  $1.2 \times 10^{-3}$ (black), $4 \times 10^{-4}$ (grey).
}
\label{fig:int}
\end{figure}

The features of the power spectrum of fluctuations have 
an intuitive and rather simple dependence on the parameter choices for the inflationary model.

\begin{enumerate}

\item
The mass hierarchy $m_{\psi}/m_{\chi}$ dictates the difference in the energy levels of the two inflationary stages.
For a larger hierarchy we expect a sharper drop from the 
plateau of the first stage to the much lower plateau of the
 second stage. This generates more numerous and also sharper turns in field space during the intermediate stage, which lasts until
 the original energy density is depleted through the effective
 friction term in the evolution equations. Ultimately, this leads to a larger enhancement of the power spectrum.
 This behaviour is apparent in the 
 left panel of Fig.  \ref{fig:int}.

\item
The interaction term controls the steepness of the walls 
around the valley in which the initial part of the field trajectory takes place. The convexity of the potential is
determined by $c_w \psi^2$, inducing a large mass term for
the $\chi$ fluctuations that act as the isocurvature modes
during this stage. As a result, these modes do not become
significant and the spectrum around the CMB scales is 
typical of single-field inflation. 
The walls around $\chi_0$ are lowered at $\psi=0$, 
allowing the evolution to continue along the $\chi$ direction 
eventually. Large values of $c_w$ reduce the size of the window
towards the second valley of the potential, so that
the fields get trapped for a longer period around the
point $(\chi_0,0)$. This results in the evolution depicted
in the right panel of Fig. \ref{fig:traj} with a large number of turns. Both curvature and isocurvature modes are activated
during this stage, resulting in the enhancement of the power
spectrum. This behaviour is apparent in the 
right panel of Fig.  \ref{fig:int}, which demonstrates that a
 larger value of $c_w$ increases the enhancement
of the spectrum.

\item
The value of 
$\psi_0$, which can be interpreted as denoting the point at which
the fields settle on the first valley of the potential after
some initial evolution,
determines the duration of the first 
inflationary stage. Thus, it fixes the number of efoldings 
$\Delta N_1=N_{\rm p}-N_*$ between the time $N_*$ that the CMB 
scales cross the horizon and the time $N_{\rm p}$ of the
transition to the second inflationary stage. 
 Similarly, $\chi_0$ controls the duration of the second stage $\Delta N_2=N_{\rm end}-N_{\rm p}$, and thus the total number of efoldings.
 
\end{enumerate}

\section{The relevance of the mechanism for general two-field models}\label{sec:gen}

Our discussion of the  mechanism of enhancement was carried out with the 
simple potential \eqref{equ:xsq}. The main ingredients of our
scenario are the two inflationary stages taking place on two plateaux 
along two straight directions of the potential, and an intermediate
stage during which the fields are trapped around the
intersection of the two directions. The 
mass terms, 
which are parts of any two-field potential, 
are the simplest means to control the ratio 
of the energy scales characterizing the two plateaux. 
A large hierarchy between them implies that
the more massive field ($\psi$ in our scenario) has large
kinetic energy after the end of the first inflationary stage,
of the order of the vacuum energy during this
stage. If the vacuum energy of the second inflationary stage is
much lower than that of the first one,
it takes long for the kinetic energy to be depleted by the effective friction term.
Thus the system stays in the intermediate stage for
an extended period of time. The numerous turns that take place
before the slow-roll evolution of the lighter field ($\chi$ in our scenario) sets in lead to a localized enhancement of the
power spectrum.

We expect that this type of enhancement is generically 
present in two- and 
multi-field inflation, and can arise for a broad range of potential shapes, 
including different plateau structures and interaction terms.
The essential requirement is a hierarchy between the energy scales of the two 
inflationary stages, together with parameters that allow for a transition 
regime with repeated turns in the background trajectory.
The model considered above can be readily reformulated or generalized, as long as the above essential ingredients are present. For example, additional 
higher-order terms or deformations can be included in any field direction. Such modifications of the potential during the first stage may be necessary in order to 
adjust the observables at CMB scales, such as the spectral index, to
the experimentally deduced values, without spoiling the small-scale enhancement of the spectrum.

More generally, the mechanism can be viewed as a consequence of the transient 
excitation of entropic fluctuations during the transition stage. Successive 
turns efficiently transfer power from isocurvature to curvature modes, leading 
to a localized enhancement of ${\cal P}_{\cal R}(k)$ without the need for 
fine-tuned features. From this perspective, the detailed functional form of 
the potential is secondary. 
Assisted enhancement should therefore be a generic possibility in a 
wide class of multi-field models motivated by effective field theory or 
supergravity.
For example, we have verified that similar behaviour occurs for potentials 
involving hyperbolic tangent terms, such as
\begin{equation}\label{equ:tanh}
	V(\chi,\psi)=c_\chi\tanh^2(\chi-\chi_1)+c_\psi\tanh^2(\psi)
	+c_w\psi^2\tanh^2(\chi-\chi_0)\, .
\end{equation}
Such structures commonly appear in supergravity constructions 
\cite{Lahanas:1986uc,Ellis:2013xoa} and cosmological attractor models 
\cite{Kallosh:2013daa,Kallosh:2013maa}, while couplings of this type also arise 
in inflationary scenarios with non-minimal interactions with gravity 
\cite{Garcia-Bellido:2011kqb,Kaiser:2013sna,Ema:2017rqn}.

\section{Gravitational waves and primordial black holes} \label{sec:PBHGW}

Significant effort is currently being devoted to the detection of gravitational 
waves across a wide range of frequencies. Ground-based interferometers such as 
LIGO-Virgo-KAGRA (LVK) have established GW astronomy through 
compact binary mergers, 
while pulsar timing array (PTA) collaborations, including NANOGrav and the IPTA, 
have reported strong evidence for a nanohertz stochastic gravitational wave 
background with Hellings--Downs-like correlations 
\cite{NANOGrav:2023gor,NANOGrav:2023hvm,EPTA:2023fyk,Reardon:2023gzh,Xu:2023wog}. 
Future space-based missions such as LISA will probe the complementary millihertz 
band, together with next generation terrestrial detectors bridging PTA and 
ground-based sensitivities \cite{LISA:2017pwj,ET:2025xjr}.

While many GW signals originate from late-time astrophysical sources, an 
especially intriguing possibility is a primordial contribution from the early 
Universe \cite{Caprini:2018mtu, Guzzetti:2016mkm}. Scalar perturbations generated 
during inflation inevitably source gravitational waves at second order, but in 
minimal slow-roll models the resulting signal is unobservably small.  
However, inflationary scenarios with 
localized enhancements of the curvature power spectrum, such as those produced 
by the sequences of turns in multi-field trajectories discussed here, can generate a strong 
stochastic background upon horizon reentry, potentially accompanied by PBH 
formation \cite{Young:2014ana,Ozsoy2023Review,Figueroa:2023zhu, Balaji:2022dbi, Balaji:2023ehk}.

The scale of the enhancement can be parameterized by the number of efoldings after 
the CMB pivot scale $k_*=0.05\,{\rm Mpc}^{-1}$ exits the horizon. A peak at 
$k_{\rm p}\simeq 0.38\times 10^{P}\,{\rm Mpc}^{-1}$ corresponds to
\begin{equation}
N_{\rm p}-N_*\simeq \ln\!\left(\frac{k_{\rm p}}{k_*}\right)\simeq 2+2.3\,P \, .
\end{equation}
Features appearing $\sim 16$ efoldings after CMB horizon exit lead to scalar-induced gravitational waves (SIGWs) in the 
nanohertz PTA band, while $\sim 30$ efoldings correspond to millihertz frequencies 
targeted by LISA. The associated mapping between frequency, comoving scale, and 
characteristic PBH mass is approximately
\begin{equation}
k \simeq 6\times10^{5}\left(\frac{f}{10^{-9}\,{\rm Hz}}\right){\rm Mpc}^{-1},
\qquad
M_{\rm H} \simeq 33\left(\frac{10^{-9}\,{\rm Hz}}{f}\right)^2 M_\odot \, .
\end{equation}
Thus, PTA frequencies are naturally linked to solar-mass PBHs, whereas LISA 
frequencies correspond to sublunar or asteroid-mass PBHs, mass windows that are 
among the least constrained and may allow PBHs to constitute a significant 
fraction of the dark matter \cite{Carr:2021bzv,Carr:2023tpt}.

\subsection{Non-Gaussianities}

The enhancement mechanism discussed in this work relies on repeated turns of the 
inflationary trajectory in field space, which generically induce superhorizon 
evolution of the comoving curvature perturbation ${\cal R}$ through its coupling 
to entropic  modes.
Periods of rapid turning not only amplify the scalar power spectrum but also 
inevitably generate non-Gaussian curvature perturbations via nonlinear 
entropic--adiabatic interactions. In the effective single-field description, 
applicable during strong turns, the adiabatic mode propagates with a reduced 
sound speed, leading to equilateral-type non-Gaussianity with amplitude 
$f_{\rm NL}^{\rm equil}\sim\mathcal{O}(c_s^{-2})$ 
\cite{Achucarro:2012sm,Chen:2010xka}.

In the scenarios considered here, the turning dynamics is localized in efolding 
time during the brief transition between two inflationary stages. Consequently, 
both the enhancement of the curvature power spectrum ${\cal P}_{\mathcal{R}}(k)$ 
and the associated non-Gaussianities are strongly scale dependent, affecting 
only modes exiting the horizon near the peak scale $k_{\rm p}$. We expect that CMB-scale modes 
remain essentially unaffected and continue to satisfy current observational 
constraints \cite{Planck:2019kim}.

Despite this localization, non-Gaussian corrections enter exponentially in the 
PBH formation probability. As a result, even moderate values 
$f_{\rm NL}^{\rm loc}=\mathcal{O}(1)$ can change the predicted PBH abundance by 
several orders of magnitude \cite{Byrnes:2012yx,Young:2013oia}. Non-Gaussianity 
can also modify the amplitude and spectral shape of SIGWs \cite{Cai:2018dig}. A dedicated treatment of these effects is therefore 
essential for precision PBH predictions, but lies beyond the scope of the 
present work.

\subsection{Primordial Black Holes and Observational Constraints}

Assuming Gaussian primordial perturbations and approximately spherical collapse, 
the fraction of the Universe collapsing into PBHs during radiation domination is 
\begin{align}
\label{brad}
\beta(M)
&=\int_{\delta_c}^{\infty}\frac{d\delta}{\sqrt{2\pi\sigma^2(M)}}
\exp\!\left(-\frac{\delta^2}{2\sigma^2(M)}\right)
\simeq \frac{1}{2}\,\mathrm{erfc}\!\left(\frac{\delta_c}{\sqrt{2}\,\sigma(M)}\right),
\end{align}
where $\sigma(M)$ is the variance of the density contrast and $\delta_c$ is the 
collapse threshold. Since $\beta(M)$ depends exponentially on $\delta_c$, PBH 
abundance predictions are extremely sensitive to its value. The threshold is not 
universal \cite{Carr:1975qj,Niemeyer:1997mt,Musco:2012au,Harada:2013epa} and is 
typically treated as an effective parameter in the range $\delta_c\sim0.3$--$0.5$.

The present-day PBH dark matter fraction,
$f_{\rm PBH}(M)\equiv\Omega_{\rm PBH}(M)/\Omega_{\rm DM}$, is proportional to 
$\beta(M)$. Consequently, the predicted abundance is also highly sensitive to 
primordial non-Gaussianities, since PBH formation probes the extreme tail of the 
probability distribution \cite{Byrnes:2012yx,Young:2013oia,Franciolini:2018vbk, Atal:2018neu, DeLuca:2019qsy, Firouzjahi:2023xke}.

PBHs may alternatively form during an early matter-dominated era, as could occur 
for sufficiently late reheating. In this case the collapse fraction scales 
polynomially with the variance \cite{Harada:2016mhb,Harada:2017fjm},
\begin{equation}\label{bmat}
\beta(\sigma)=0.056\,\sigma^5,
\end{equation}
implying a significantly reduced sensitivity to non-Gaussian corrections.

Across most of the PBH mass spectrum, stringent bounds constrain the allowed 
fraction of dark matter in PBHs \cite{Carr:2020xqk,Carr:2020gox,Green:2024bam}. 
The weakest constraints occur in the window 
$M_{\rm PBH}\sim10^{-15}$--$10^{-10}\,M_\odot$, which is particularly relevant 
for LISA through the associated SIGW signal. By shifting the enhancement scale 
of ${\cal P}_{\mathcal{R}}(k)$, PBHs can also form in other mass ranges, 
including the stellar-mass window relevant for LIGO-Virgo-KAGRA mergers.

In Fig.~\ref{fig:GWPBH} we show the resulting PBH abundance for the benchmark 
potential \eqref{equ:xsq}. During radiation domination the predictions are non-robust because of their sensitivity to $\delta_c$ and non-Gaussianities, 
whereas PBH formation in an early matter-dominated era is considerably more 
stable. By comparison, the associated SIGW spectrum, discussed below,  retains more direct 
information about the shape of the primordial scalar power spectrum.

\subsection{Scalar-induced stochastic gravitational waves}
\label{sec:SIGW}

An approximately isotropic stochastic gravitational-wave background is 
conventionally characterized by the dimensionless energy density per logarithmic 
frequency interval,
\begin{equation}
\Omega_{\rm GW}(f)\equiv 
\frac{1}{\rho_{\rm tot}}\frac{d\rho_{\rm GW}}{d\ln f},
\qquad
\rho_{\rm tot}=\frac{3H_0^2}{8\pi G}.
\end{equation}

In our scenario, strong scalar-induced gravitational waves are mainly produced 
when the enhanced scalar perturbations reenter the Hubble horizon. During 
radiation domination, the observable induced tensor spectrum can be computed within 
second-order cosmological perturbation theory 
\cite{Ananda:2006af,Baumann:2007zm, Espinosa:2018eve, Kohri:2018awv, Domenech:2021ztg, Domenech:2025ccu}. The oscillation-averaged 
tensor power spectrum takes the form \cite{Kohri:2018awv}
\begin{equation}
\label{tensorPSD2}
\overline{\mathcal{P}_{h}(\tau,k)}
=\int_{0}^{\infty} dt \int_{-1}^{1} ds\ 
\mathcal{T}(s,t,\tau,k)\,
{\cal P}_{\cal R}\!\left(\frac{t+s+1}{2}k\right)
{\cal P}_{\cal R}\!\left(\frac{t-s+1}{2}k\right),
\end{equation}
where $\mathcal{T}$ is a known kernel \cite{Kohri:2018awv, Domenech:2021ztg}. 
The present-day GW energy density is then approximately
\begin{equation}
\label{OmegaIGWc}
\Omega_{\rm GW}(t_0,k)
\simeq g_*^{-1/3}\,\frac{\Omega_{\gamma,0}}{24}
\left(\frac{k}{aH}\right)^2
\overline{{\cal P}_h(k)} \, .
\end{equation}

If horizon reentry occurs instead during an early matter-dominated era, density 
perturbations may become nonlinear prior to reheating, leading to qualitatively 
different GW spectra. In this case a
dominant contribution can be estimated from gravitational radiation emitted 
during collapse and halo formation
by the quadrupole moment $\tilde{Q}_{ij}$ \cite{Dalianis:2020gup,Dalianis:2024kjr}:
\begin{align}
\Omega_\text{GW}(t_0,\omega_0)=\frac{1}{\rho_\text{tot}(t_0)}
\int_{\cal S} d\alpha\,d\beta\,d\gamma\,
\frac{1}{1+z_{\rm e}}\frac{1}{V_k}\frac{4\pi G}{5c^5}
\sum_{ij}\!\left|\tilde{Q}_{ij}(\omega)\right|^2 \omega^7\,
{\cal F}_\text{D}(\alpha,\beta,\gamma,\sigma)\,,
\label{result_Om}
\end{align}
where ${\cal S}$ denotes the region leading to consistent nonspherical collapse,
$z_{\rm e}$ is the emission redshift, $V_k=4\pi k^{-3}/3$ is the comoving  
horizon-entry volume, and  ${\cal F}_\text{D}$ is a probability density function.

Primordial non-Gaussianity can modulate both the amplitude and spectral shape of 
the induced signal \cite{Cai:2018dig}. However, unlike PBH formation, which is 
exponentially sensitive to the tail of the probability distribution, the SIGW 
amplitude is mainly determined by the scalar power spectrum. SIGWs therefore 
provide a comparatively robust probe of enhanced primordial fluctuations on 
small scales.

Large scalar perturbations associated with PBH production can generate signals 
as large as $h^2\Omega_{\rm GW}(f_{\rm p})\gtrsim10^{-14}$--$10^{-12}$, bringing 
them within reach of PTA experiments and future detectors such as LISA. In 
particular, multi-peak structures in ${\cal P}_{\cal R}(k)$ are partially 
inherited by the GW spectrum, providing an observational window into the 
underlying inflationary dynamics.

\begin{figure}[t]
  \centering
  \begin{subfigure}[t]{0.48\textwidth}
    \centering
    \includegraphics[width=\textwidth]{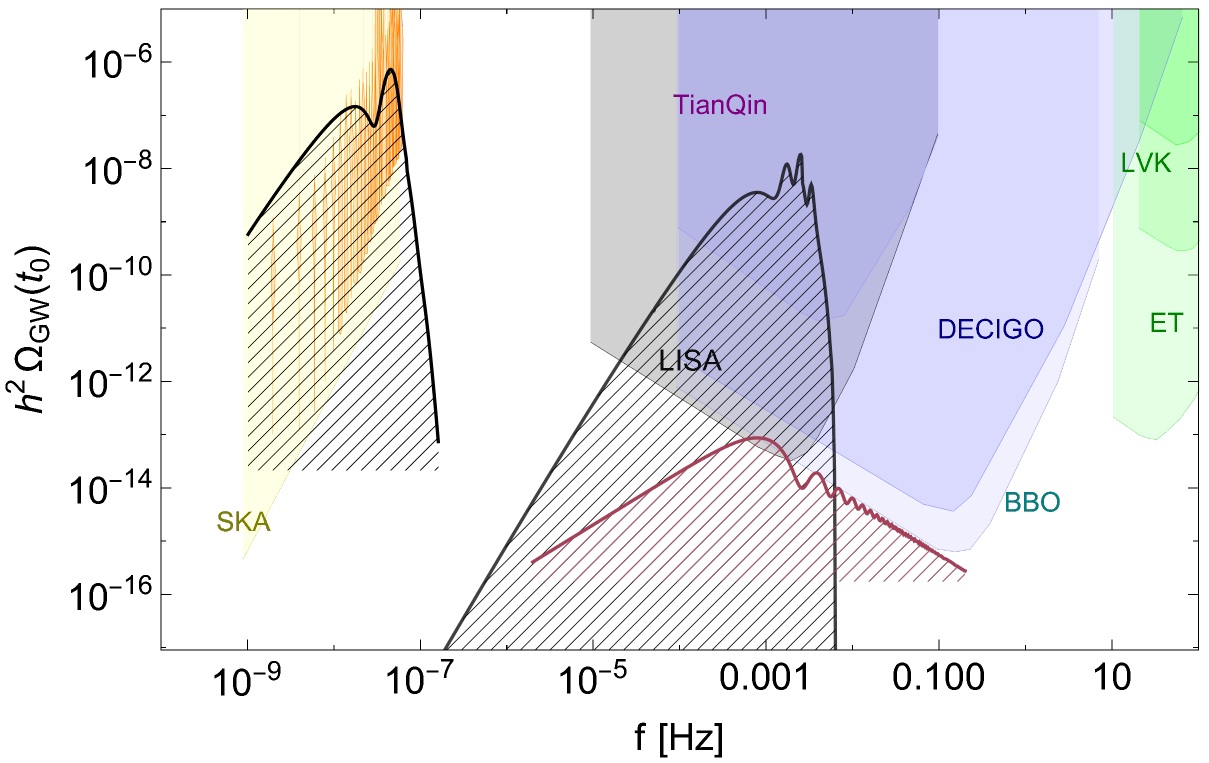}
    \label{fig:twopanel:left}
  \end{subfigure}
  \hfill
  \begin{subfigure}[t]{0.48\textwidth}
    \centering
    \includegraphics[width=\textwidth]{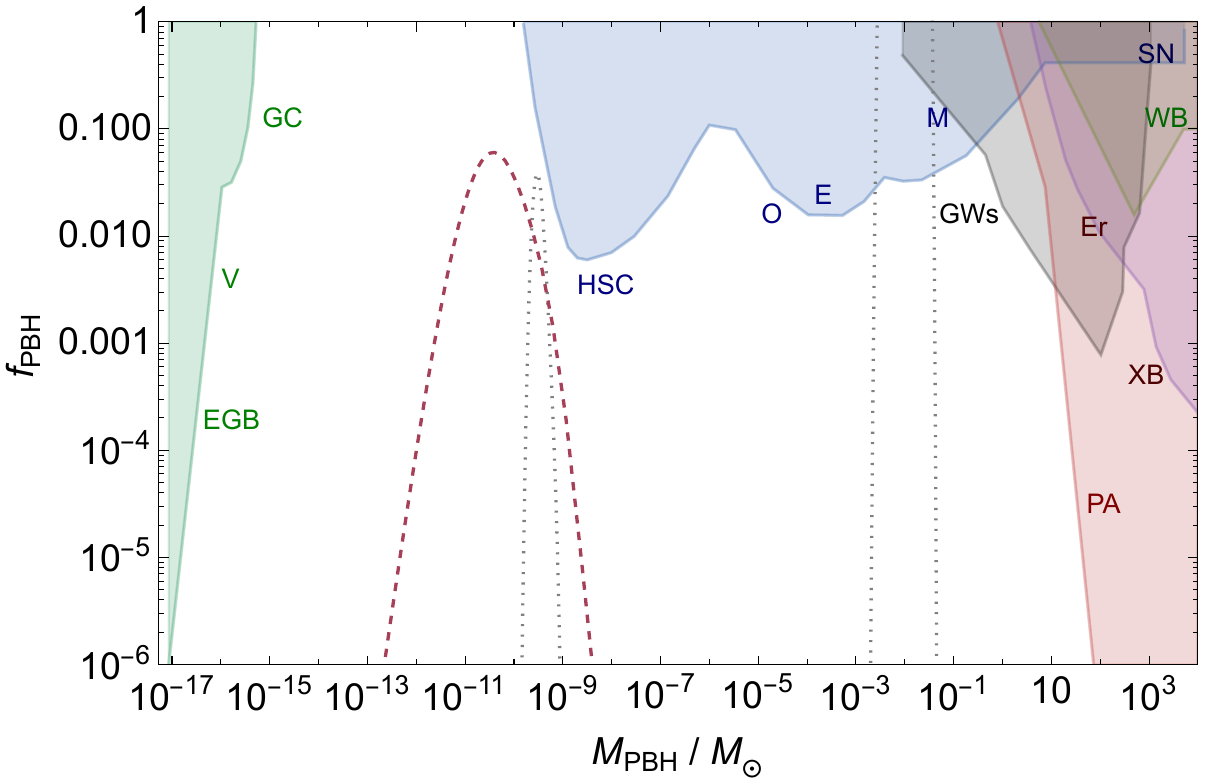}
  \end{subfigure}
  \caption{ { Left panel}: Spectrum of scalar-induced GWs in our two-field scenario. The gray (dark-red) curve shows the signal produced during 
radiation domination (an early matter-dominated era), up to 
non-Gaussian corrections. Orange lines indicate PTA posteriors from NANOGrav, 
EPTA, and InPTA \cite{NANOGrav:2023hvm,EPTA:2023fyk}.  
{ Right panel}: The corresponding PBH fractional abundance. The dotted gray 
(dashed dark-red) curve denotes PBHs formed during radiation domination (early 
matter domination). Shaded regions show current observational constraints. 
These PBH predictions are only indicative because of their sensitivity to 
non-Gaussian effects.}

\label{fig:GWPBH}
\end{figure}

Representative SIGW spectra for the benchmark potential \eqref{equ:xsq} are shown 
in Fig.~\ref{fig:GWPBH}, including both radiation- and matter-dominated reentry 
scenarios, and compared with the frequency ranges probed by LISA and current PTA data 
\cite{NANOGrav:2023gor,EPTA:2023fyk}. The parameters of the
corresponding models are given in Table \ref{tab:params}.

\section{Conclusions and outlook}
 \label{conclusions}

In this work we have discussed a simple and generic mechanism to generate a strong enhancement of the primordial curvature power spectrum in two-field models of inflation.
The key ingredient is a hierarchy of energy scales that gives rise to a two-stage inflationary evolution.
During the transition between the two stages, the inflationary trajectory undergoes a sequence of sharp turns in field space, leading to an efficient transfer of power from entropic to adiabatic perturbations.
The mechanism does not rely on fine-tuned inflection points, ultra-slow-roll phases, or specially engineered valleys in the potential. Instead, 
it can be realized within a simple setup consisting of 
quadratic potentials for each field, supplemented by a minimal interaction term. 
This is sufficient to generate a strong enhancement of the curvature power 
spectrum ${\cal P}_{\mathcal{R}}(k)$ on small scales, while leaving the CMB 
scales unaffected by the amplification and free from significant non-Gaussian 
contributions.

The effects of turns and sharp features on the curvature perturbations can be understood in terms of localized departures from slow-roll, which act as transient sources for the curvature perturbation ${\cal R}$, as reviewed in
the introduction. When such events occur repeatedly and in close succession, their contributions can interfere constructively, leading to a resonant enhancement of the power spectrum. 
The present work provides a very simple dynamical realization of this mechanism. The repeated turns in field space naturally emerge from the oscillatory relaxation of one field during the transition between two stages of inflation, generating a sequence of localized pulses. The turns do not have to be particularly sharp or enhanced by curvature in field space or kinetic factors.
% They emerge as a natural relaxation trajectory. 
%Furthermore, the localization of the feature is achieved by the Hubble friction, and substitutes the ad-hoc window function of earlier models \cite{Katsis:2025eed}. 
The analytic intuition developed in the pulse approximation applies here, with the total enhancement controlled by the parameter $W$,
defined in \eqref{equ:cond}, and the duration $\Delta N$ of the turning phase, as discussed in Refs.~\cite{Boutivas:2022qtl,Fumagalli:2020adf}.

We illustrated the mechanism using a minimal two-field setup with canonical kinetic terms and a simple interaction potential \eqref{equ:xsq}.
The first stage of inflation proceeds at a higher energy scale and sets the CMB normalization, while the second stage occurs at a lower scale.
As the system relaxes from the first to the second stage, oscillatory motion in the heavier field induces repeated turns of the background trajectory,  see Fig. \ref{fig:traj}.
These turns act as transient sources for curvature perturbations, producing a localized amplification of the power spectrum over a narrow range of scales.  Only modes that exit the horizon during the brief transition period are amplified, while both earlier and later modes remain unaffected.
This naturally leads to a peaked primordial power spectrum, whose position and amplitude are controlled by the duration of the transition, the mass hierarchy, and the strength of the interaction between the fields.
As a result, the curvature power spectrum can grow by several orders of magnitude at short scales while remaining consistent with the CMB at large scales.

It must be pointed out that our simple setup employs only 
quadratic terms in the slow-roll phases. 
The effect on the scalar spectral index and the tensor-to-scalar ratio
can be phenomenologically important. 
Modifications of the potential along the plateau of the
first inflationary stage may be necessary in order to 
adjust such observables to
the experimentally deduced values.
However, the purpose of this work was to demonstrate a simple and generic  mechanism for power-spectrum enhancement, rather than to construct a detailed inflationary model with perfect agreement with observations. We expect that this type of enhancement is broadly applicable in two- and 
multi-field inflation, and can arise for a broad range of potential shapes.
The mass terms included in the potential  \eqref{equ:xsq}, which set the energy scales for the two inflationary stages, will appear in a
small-field expansion of more general potentials, such as those
in the earlier works \cite{Pi:2017gih,He:2018gyf,Wang:2024vfv,Kim:2025dyi,Wang:2025dbj}.

 %In fact the turns themselves do not even have to be elaborately sharp or specially crafted, the simple trajectory of relaxing towards the minimum potential in one field direction is already sufficient, for the right hierarchy of scales between the two plataux.

The enhanced small-scale curvature perturbations have direct implications for PBH  formation.
Modes that reenter the horizon during radiation domination with sufficiently large amplitude can collapse into PBHs. 
 We mention that our mechanism is expected to generate significant non-Gaussianity at short scales, which can
substantially modify PBH abundance estimates based on
%and modify SIGW predictions 
the simplest Gaussian treatment. A more detailed computation of the PBH abundance including non-Gaussian corrections, which are generically expected in turning trajectories, would refine the phenomenological predictions presented in this work. In any case, the possibility of observing non-Gaussianities of primordial fluctuations is very exciting, as it might provide information on the field content during inflation.
  
An unavoidable and complementary prediction of this framework is the generation of a stochastic GW background induced at second order by scalar perturbations.
The resulting  scalar-induced gravitational wave (SIGW) background may be detectable by current and future GW experiments. It would be very interesting
to identify specific features of the SIGW background that reflect the 
oscillatory patterns in the adiabatic spectrum, along the lines
of Ref. \cite{Fumagalli:2020nvq}. 
Such features are visible in the left panel of Fig. \ref{fig:GWPBH}.
However, it is not straightforward to correlate them with the 
characteristic energy scales
of the potential \eqref{equ:xsq}. The complicated pattern in the
right panel of Fig. \ref{fig:traj} results from oscillations 
of the fields $\psi$ and $\delta\chi=\chi-\chi_0$ around zero. The 
presence of an interaction term $\sim \psi^2 \delta\chi^2$ and
the damping of oscillations by the effective Hubble friction term 
shift the characteristic frequencies. Moreover, the numerous 
oscillations
before the system settles on the second inflationary trajectory 
lead to very complicated interference patterns in the 
adiabatic spectrum, in agreement with what was observed in 
Ref. \cite{Boutivas:2022qtl}. The difficult task of identifying 
quantitative signatures of these 
patterns in the SIGW background will be the subject of future work.

%The localized nature of the power-spectrum enhancement, the possible peak structure inherited from the turning dynamics, and the correlation between PBH masses, GW frequencies and non-Gaussianities offer distinctive signatures that can be used to discriminate this scenario from single-field ultra-slow-roll or inflection-point models.  
%The mechanism presented here highlights 
%how seemingly generic features of multi-field dynamics can leave striking observable imprints on small-scale cosmological structures.

\end{document}